\documentclass[twocolumn,aps,prl,groupedaddress,showpacs]{revtex4-1}
\usepackage{graphicx} 
\usepackage{amsmath}
\usepackage{amssymb}
\usepackage{bm}   
\usepackage{grffile}
\usepackage{subfigure}

\begin{document}

\title{Optimal Network Modularity for Information Diffusion}%

\author{Azadeh Nematzadeh}
\affiliation{School of Informatics and Computing, Indiana University, Bloomington, IN 47408 USA}

\author{Emilio Ferrara}
\affiliation{School of Informatics and Computing, Indiana University, Bloomington, IN 47408 USA}

\author{Alessandro Flammini}
\affiliation{School of Informatics and Computing, Indiana University, Bloomington, IN 47408 USA}

\author{Yong-Yeol Ahn}\email[Corresponding author: ]{yyahn@indiana.edu}
\affiliation{School of Informatics and Computing, Indiana University, Bloomington, IN 47408 USA}

\date{\today}

\begin{abstract}

We investigate the impact of community structure on information diffusion with
the linear threshold model. Our results demonstrate that modular structure may
have counter-intuitive effects on information diffusion when social
reinforcement is present.  We show that strong communities can facilitate
global diffusion by enhancing local, intra-community spreading.  Using both
analytic approaches and numerical simulations, we demonstrate the existence of
an optimal network modularity, where global diffusion require the minimal
number of early adopters. 

\end{abstract}


\maketitle

The study of information diffusion---fads, innovations, collective actions,
viral memes---is relevant to a number of disciplines, including mathematical,
physical and social sciences, communication, marketing and
economics~\cite{ryan1943diffusion, granovetter1978threshold,
rogers2010diffusion, Gruhl2004information, bakshy2012role, weng2013virality}.
The most common approach is to focus on the affinities between information
diffusion and infectious diseases spreading~\cite{goffman64, Daley1964rumour}:
a piece of information can travel from one individual to another through social
contacts and the ``infected'' individuals can, in turn, propagate the
information to others, possibly generating a large-scale diffusion event
similar to an epidemic outbreak~\cite{bailey75, anderson92epidemics}.  In
addition to classical epidemic models, two main types of information diffusion
models have been proposed: the \textit{independent cascade model}, which was
initially adopted to study the dynamics of viral
marketing~\cite{goldenberg2001talk, saito2008prediction, domingos2001mining,
richardson2002mining, leskovec2007dynamics}, describes information diffusion as
a branching process; the \textit{threshold model},  originally proposed to
study collective social behavior~\cite{schelling1971dynamic,
granovetter1978threshold, watts2002simple, krapivsky2011fads}, incorporates the
idea of `social reinforcement' by assuming that each adoption requires a
certain number of exposures. Although it is not yet fully understood how the
microscopic mechanisms underlying information diffusion differ from those in
epidemic spreading, it has been pointed out that social reinforcement could be
a crucial one: unlike epidemic spreading, where each exposure acts
independently, social reinforcement provisions that each additional exposure to
a piece of information sensibly increases the probability of its
adoption~\cite{backstrom2006group, Romero2011differences, centola2010spread}.

Since information spreads through social contacts, the structure of the
underlying social network is a crucial ingredient in modeling information
diffusion.  The role of hubs and degree distribution have been studied
extensively due to their critical role in epidemic
spreading~\cite{Pastor2001epidemic, albert2002statistical, newmanbook}.
Another obvious network feature that has implications on information diffusion
is the presence of a modular structure. Several studies investigated the role
of communities in information diffusion~\cite{onnela2007structure,
gleeson2008cascades, lambiotte2009knowledge, ikeda2010cascade,
hackett2011cascades, chung2013epidemic}, mostly ignoring the effect of social
reinforcement.

Epidemic spreading is hindered by the presence of communities or modular
structure, since this helps confining the epidemics in the community of
origin~\cite{onnela2007structure, wu2008community}. This may naturally lead to
the expectation that the same is true for information diffusion, given the
similar  approaches used in modeling epidemic and information diffusion.
However, recent empirical work suggested that modular structure may,
counterintuitively, \emph{facilitate} information
diffusion~\cite{centola2010spread}.  Other studies also proposed that
network modularity  plays a more important role in  information diffusion than
in epidemics spreading~\cite{backstrom2006group, weng2013virality,
weng-viral-2014}. These findings reinforce the need to systematically explore
how mechanisms like social reinforcement interact with the ubiquitous presence
of modular structure in the underlying network. 

In this letter we use the linear threshold model---which incorporates the
simplest form of social reinforcement---to systematically study how community
structure affects global information diffusion.  It is worth stressing that
both cooperative interactions (as those provisioned by social reinforcement)
and modular structure are common in a variety of phenomena. The results
described here could be, therefore, directly relevant in several different
areas. Examples include neural networks~\cite{gerstner2002spiking}, systems
with Ising-like dynamics evolving on a non-homogenous
substrate~\cite{chen2011optimal}, and more in general in the study of phenomena
that can be interpreted in terms of spreading.

We here expose two roles played by modular structure: \emph{enhancing local
spreading} and \emph{hindering global spreading}.  Strong communities
facilitate social reinforcement and thereby enhance local
spreading~\cite{centola2010spread, weng2013virality}; weak community structure
makes global spreading easier, because it provides more \emph{bridges} among
communities.  We show that there exists an \emph{optimal balance} between these
two effects, where community structure counterintuitively
\emph{enhances}---rather than hinders---global diffusion of information.  This
draws a parallel with the `small world' phenomenon, where the presence of
a small number of shortcuts greatly reduces the average path length of the
network while maintaining high clustering~\cite{watts1998collective}. In
information diffusion, a small number of bridges between communities allows
inter-community diffusion while maintaining intra-community diffusion. 

We adopt the linear threshold model to account for recent observations and
experiments that demonstrated the impact of social reinforcement in information
diffusion~\cite{backstrom2006group, centola2010spread, Romero2011differences,
weng2013virality}.  Let us formally define the linear threshold model first.
Consider a set of $N$ nodes (agents) connected by $M$ undirected edges.  The
state of an agent $i$ at time $t$ is described by a binary variable $s_i (t)
=\{0,1\}$, where $1$ represents the `active' state and $0$ the `inactive' one.
At time $t=0$ a fraction $\rho_0$ of randomly selected  agents, or `seeds,' is
initialized in the active state.  At each time step, every agent's state is
updated synchronously according to the following threshold rule:  \[ s_i(t+1) =
\left\{ \begin{array}{ll}
1 & \quad \text{if}\quad \theta k_i < \sum_{j \in \mathcal{N}(i)} s_j(t),\\
0 & \quad \text{otherwise},
\end{array}\right. \]
where $\theta$ is the threshold parameter, $k_i$ is the degree of node $i$, and
$\mathcal{N}(i)$ the set of $i$'s neighbors.  This rule implies that:
\textit{(i)} the dynamics is deterministic; \textit{(ii)} once a node becomes
active, it will remain so forever; and, \textit{(iii)} if $s_i(t+1) = s_i(t)$
for all nodes, then the system is in a steady state.  The linear threshold
model exhibits various critical behaviors.  For instance, there is a critical
threshold parameter at which a single active node can trigger a macroscopic
cascade~\cite{watts2002simple}; there also exists a sharp transition, at a
constant threshold parameter, from an inactive state where no diffusion occurs,
to an active state with global diffusion, triggered at a critical fraction of
initially active nodes~\cite{singh2013threshold}.  In the following,  we focus
on the latter transition based on the number of seeds and let $\theta$
constant. 

\begin{figure}
\subfigure[$\mu$=0.03]{\label{fig3:a}\includegraphics[width=0.32\columnwidth]{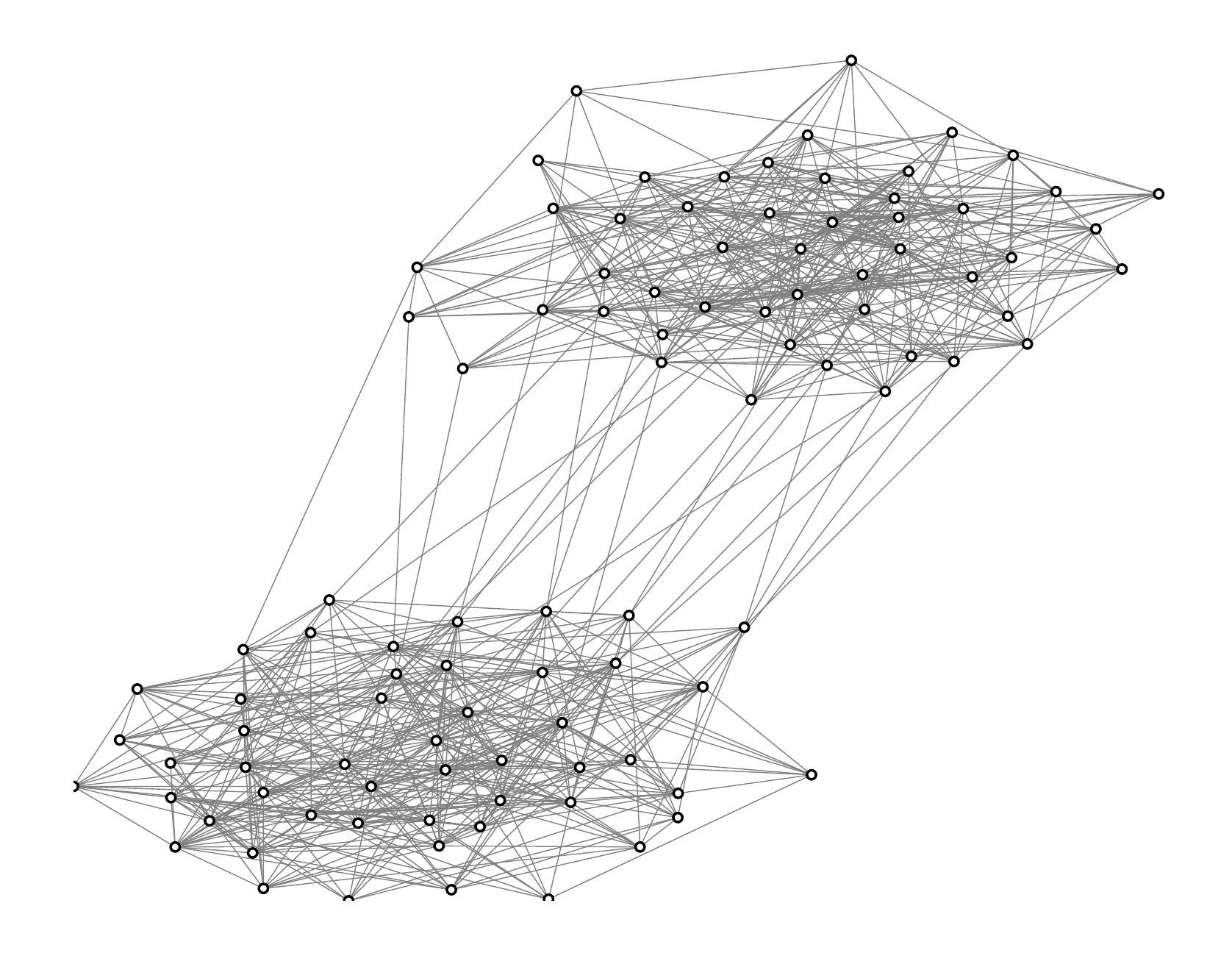}}
\subfigure[$\mu$=0.12]{\label{fig3:b}\includegraphics[width=0.32\columnwidth]{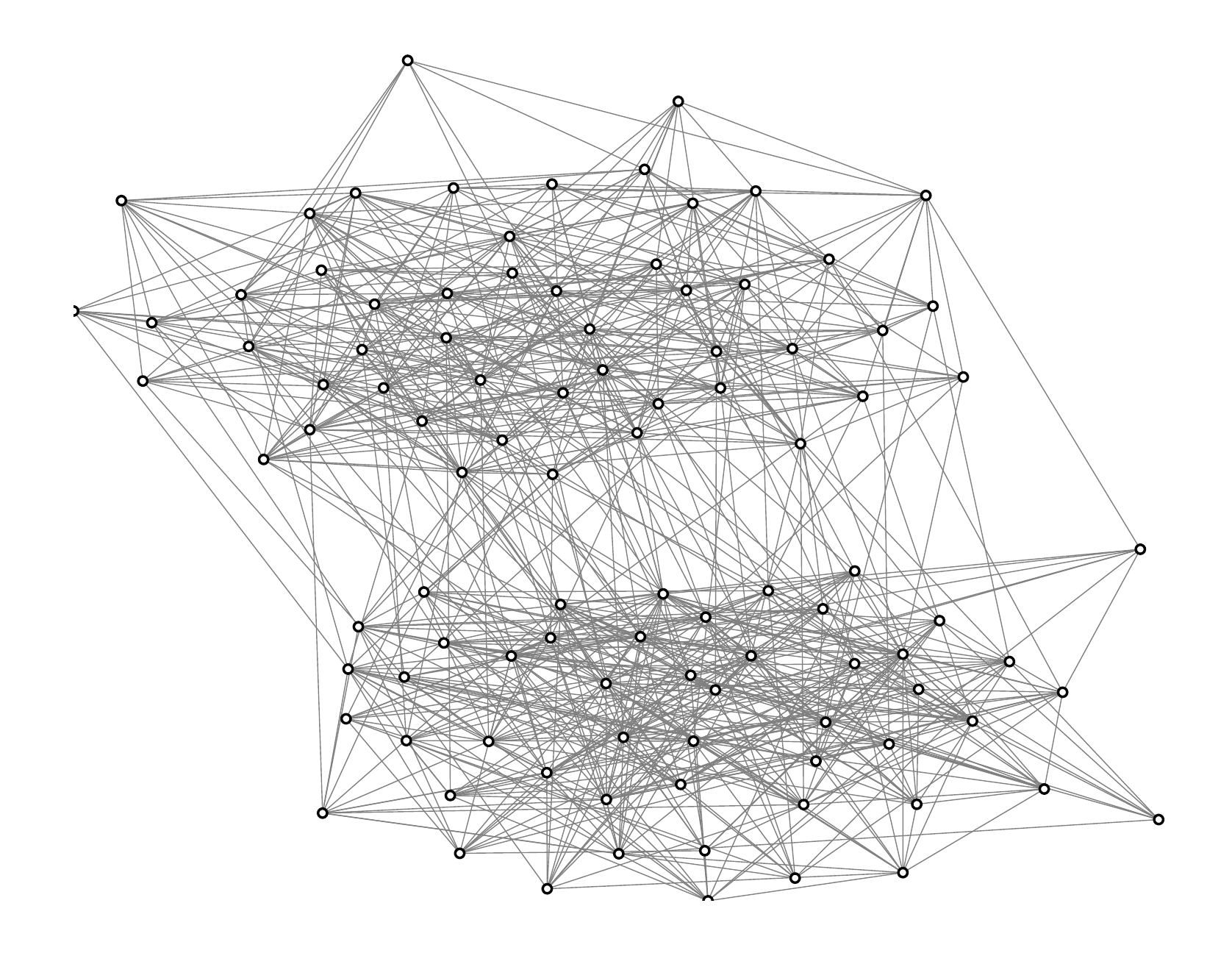}}
\subfigure[$\mu$=0.3]{\label{fig3:c}\includegraphics[width=0.32\columnwidth]{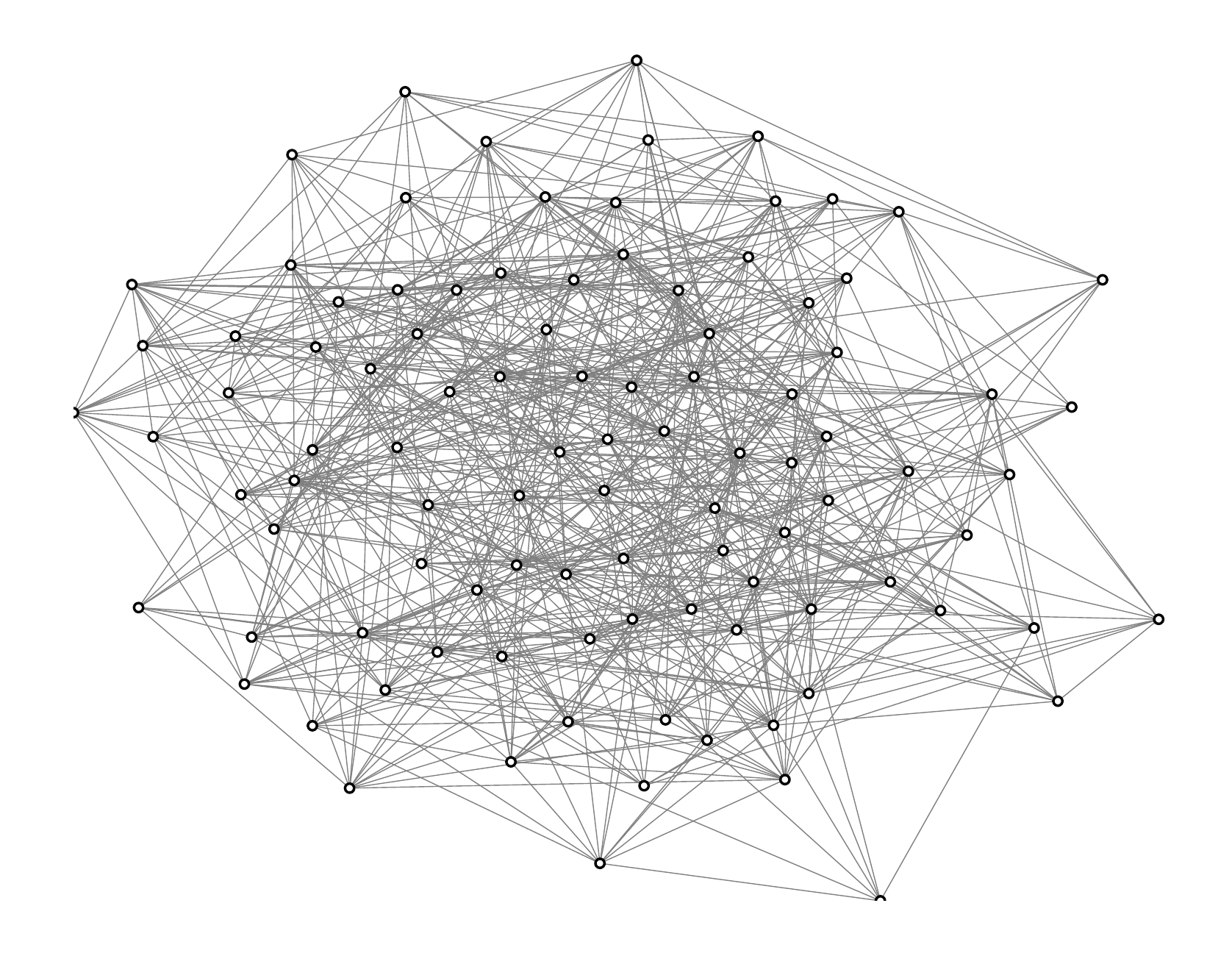}}

\caption{Example of networks with different degrees of clustering $\mu$.
Parameters values are set to $N = 100$, $M=750$, and $n = 2$.
}\label{fig:network}

\end{figure}

To systematically investigate the impact of community structure, we prepare an
ensemble of networks with two communities with varying degree of strength,
using the block-model approach~\cite{girvan2002community, lancichinetti2008benchmark,
karrer2011stochastic}.  First, half of the nodes are randomly selected and
assigned to community $A$, and the other half are assigned to community $B$.
Then, $(1-\mu)M$ links are randomly distributed among node pairs in the same
community and $\mu M$ are randomly distributed among node pairs that belong to
different communities (see Fig.~\ref{fig:network}).  The parameter $\mu$
controls the strength of the community structure: a large value of $\mu$ yields
more links between the two communities and thus a weak community structure.
Finally, we plant the seeds in $A$, assuming that the diffusion originates from
the community $A$. 

Let us introduce two analytic approaches---mean-field (MF) and tree-like (TL)
approximations---to understand the behavior of our system. We first assume
that the underlying network has a given degree distribution $p(k)$ but is
otherwise random.  We aim to compute the final density of active nodes
($\rho_\infty$) given the initial density of seeds ($\rho_0$). When there is no
community structure, using the mean-field approximation, $\rho_\infty$ can be
computed as the smallest stable solution of the equation: \begin{equation}
\rho_{\infty}=\rho_0 +(1-\rho_{0}) \sum_{k=1}^{\infty} p(k) \sum_{m=
\left\lceil \theta k \right\rceil}^{k} {k \choose m}
\rho_{\infty}^{m}(1-\rho_\infty)^{k-m}.  \label{eq:meanfield}
\end{equation} The probability that a node of degree $k$ is in the active state
at stationarity is the sum of two contributions: \textit{(i)} the probability
that the node is active  at $t=0$ ($\rho_0$); and, \textit{(ii)} the
probability that the node is not active at $t=0$ ($1-\rho_0$) but has at least
$\theta k$ active neighbors at $t = \infty$ (the second summation). The sum
over $k$ accounts for the different degrees a node may have. The equation can
be solved iteratively. 

Now let us extend Eq.~\ref{eq:meanfield} to deal with networks with communities. 
While it is easy to generalize it for arbitrary configurations of communities, here we focus on the case with two communities. 
In such a case, the equations for the fraction of active nodes $\rho^A$ (resp., $\rho^B$) in the community $A$ (resp., $B$) can be written as:
\begin{align}
\rho_{\infty}^{A(B)} &= \rho_{0}^{A(B)} + (1 - \rho_{0}^{A(B)}) \sum_{k=1}^{\infty} p(k) \nonumber \\ 
&\times \sum_{m= \left\lceil \theta k \right\rceil}^{k} {k \choose m} (q^{A(B)})^m (1-q^{A(B)})^{(k-m)}, 
\label{eq:meanfield2}
\end{align}
where $\rho_0^{A(B)}$ is the density of seeds in the community $A(B)$, and $q^{A(B)}=(1-\mu)\rho_{\infty}^{A(B)}+ \mu \rho_{\infty}^{B(A)}$ is the probability that a neighbor of a node is active, which is the sum of: \textit{(i)} the probability that the neighbor is in the same community $(1-\mu)$ and is active ($\rho_{\infty}^{A(B)}$); and,  \textit{(ii)} the probability that it is in the other $B(A)$ community ($\mu$) and is active ($\rho_{\infty}^{B(A)}$). Finally,
$\rho_{\infty} = (\rho_{\infty}^A + \rho_{\infty}^B)/2.$

A more sophisticated framework adopts the tree-like (TL) approximation~\cite{gleeson2008cascades,gleeson2007seed}. 
It approximates the underlying network with a tree of infinite depth and assumes that the nodes at level $n$ are only affected by those at level $n-1$. 
The fraction of active nodes in community $A(B)$ is computed using an auxiliary variable $y_{\infty}^{A(B)}$ obtained by the following iteration over all the levels in the tree:
\begin{align}
y_{n+1}^{A(B)}&=\rho_0^{A(B)} + (1-\rho_{0}^{A(B)}) \sum_{k}
\frac{k}{z}p(k) \nonumber \\ 
\times &\sum_{m= \left\lceil \theta k \right\rceil}^{k-1} {k-1 \choose
m} (\overline{y}_{n}^{A(B)})^{m}(1-\overline{y}_n^{A(B)})^{k-1-m},
\label{eq:q_nxt_com}
\end{align}
where $z$ is the average degree and $\overline{y}_{n}^{A(B)} = (1-\mu) y_{n}^{A(B)} + \mu y_{n}^{B(A)}$. The fraction of active nodes is given by:
\begin{align} 
\rho_{\infty}^{A(B)}&= \rho_0^{A(B)} + (1-\rho_{0}^{A(B)})
\sum_{k=0}^{\infty} p(k) \nonumber \\ \times &\sum_{m= \left\lceil \theta k \right\rceil}^{k}{k \choose
m}(y_{\infty}^{A(B)})^{m}(1-y_{\infty}^{A(B)})^{k-m}. 
\label{eq:p_nxt_com}
\end{align}

\begin{figure}
\includegraphics[width=\columnwidth, clip=true, trim=10 11 0 0]{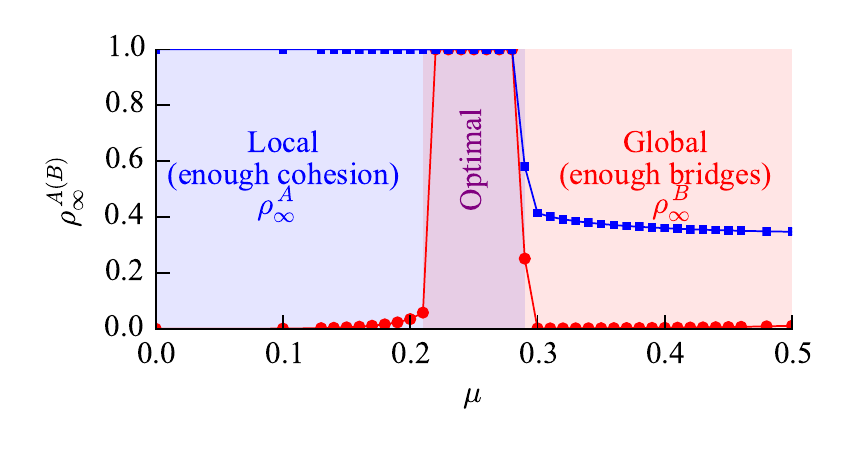}

\caption{The tradeoff between intra- and inter-community spreading.  Stronger
communities (small $\mu$) facilitate spreading within the originating community
(local) while weak communities (large $\mu$) provide bridges that allow
spreading between communities (global).  There is a range of $\mu$ values that
allow both (optimal).  The blue squares represents $\rho_\infty^A$, the final
density of active nodes in the community $A$, and the red circles represents
$\rho_\infty^B$.  The parameters for the simulation are: $\rho_0 = 0.17$,
$\theta = 0.4$, $N = 131056$, and $z = 20$. }\label{fig:optimum_mu}

\end{figure}

We now address the issue of how communities affect information diffusion by
first highlighting the trade-off due to the strength of communities.  As $\mu$
decreases,  nodes in $A$ have increasingly more neighbors in $A$.  Thus, the
number of seed nodes to which nodes in $A$ are exposed also increases because
the seeds exist only in $A$ ($\rho_0^{A} = 2\rho_0$ and $\rho_0^{B} = 0$). In
other words, strong communities \textit{enhance}  local spreading.  By
contrast, the spreading in  community $B$ is triggered entirely by the nodes in
$A$, as $\rho_0^B = 0$.  Therefore, larger $\mu$ (smaller modularity) helps the
spreading of the contagion to  community $B$.  The fact that large modularity
(smaller $\mu$) facilitates the spreading in the originating community, but
small modularity (larger $\mu$) helps inter-community spreading, raises the
following question: is there an optimal modularity that facilitates both intra-
and inter-community spreading?

Fig.~\ref{fig:optimum_mu} demonstrates that there is indeed a range of values
of $\mu$ that enables both.  In the blue range  (``local''), strong cohesion
allows intra-community spreading in the originating community $A$; in the red
range (``global''), weak modular structure allows inter-community spreading
from $A$ to $B$.  The interval where blue and red overlap (purple, ``optimal'')
provides the right amount of modularity to enable global diffusion.  Here the
modularity is large enough to initiate the local spreading and small enough to
induce inter-community spreading.  If $\mu$ is too small, the contagion cannot
propagate into $B$, even if $A$ is fully saturated,  because there are not
enough inter-community bridges.  If $\mu$ is too large, although there are
enough bridges, $\rho_\infty^B \simeq 0$ because the modularity is too small to
initiate intra-community spreading from $A$.

\begin{figure}
\includegraphics[width=\columnwidth, clip=true, trim=0 10 0 8]{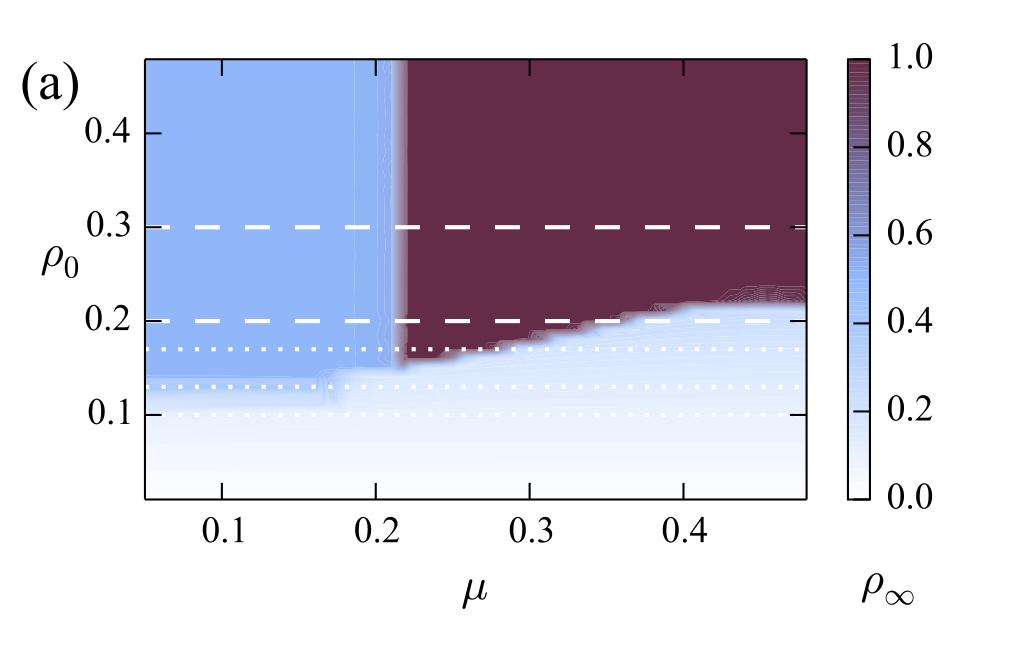}
\includegraphics[width=\columnwidth, clip=true, trim=0 10 0 0]{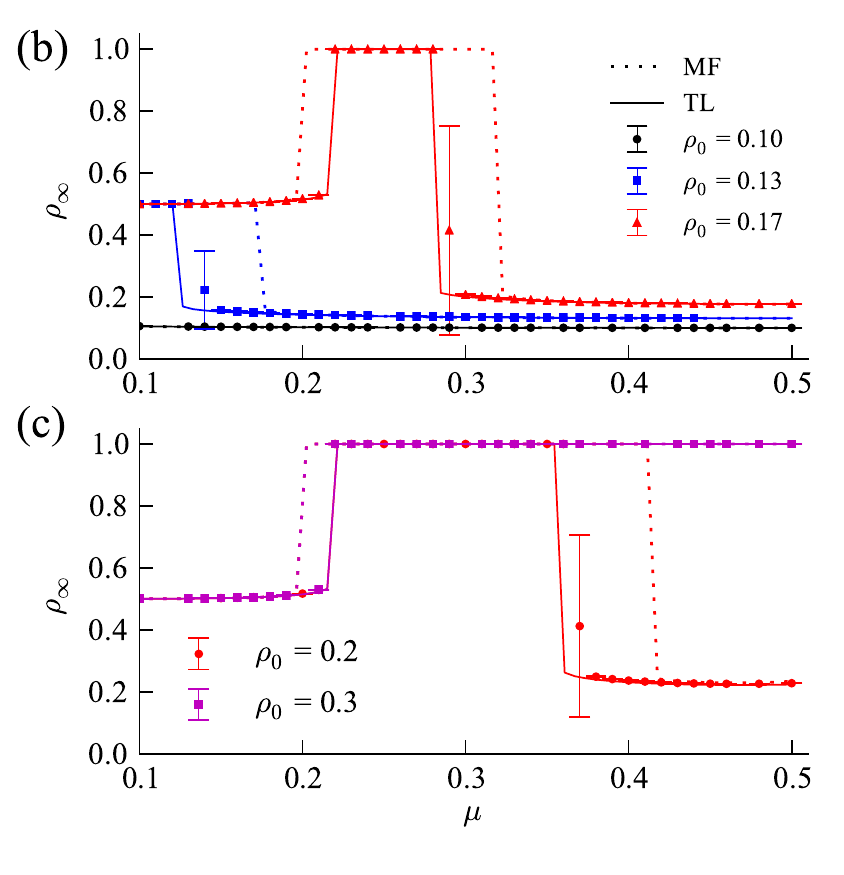}

\caption{(a) the phase diagram of threshold model in the presence of community
structures with $N = 131056$, $z=20$, and $\theta=0.4$.  There are three
phases: no diffusion (white), local diffusion that saturates the community $A$
(blue), and global diffusion (red).  The dotted and dashed lines indicate the
values of $\rho_0$ shown in (b) and (c).  (b) the cross-sections of the phase
diagram (dotted lines in (a)). TL (solid lines) shows excellent agreements with
the simulation while MF (dotted lines) overestimate the possibility of global
diffusion.  (c) the cross-sections represented in dashed lines in (a).  }
\label{fig:details}

\end{figure}

Let us analyze the issue more into detail. 
Fig.~\ref{fig:details} summarizes our results, derived analytically by MF and TL approximations, and by numerical simulations. 
In our numerical simulations, we compute the mean of $\rho_{\infty}$ across 1,000 runs of the model, each assuming a different realization of the network and of the seed nodes.  
We fix the threshold ($\theta=0.4$) throughout all simulations. We discuss the effect of various values of the threshold and other parameters, including number of communities and more general degree distributions in the supplementary material.

Fig.~\ref{fig:details}~(a) shows the phase diagram with three phases: no diffusion (white), local diffusion (blue), and global diffusion (red). 
As expected, a cross-section for $\mu=const.$ shows that $\rho_{\infty}$ is an increasing function of $\rho_0$. 
The system undergoes a sharp transition for a broad range of values of $\mu$, including the case in which communities are absent ($\mu = 1/2$)~\cite{singh2013threshold}. 
The behavior of $\rho_{\infty}$ as a function of $\mu$ is more interesting, in that it exhibits qualitatively different patterns depending on $\rho_0$.  

Fig.~\ref{fig:details}~(b,c) illustrates a set of possible scenarios, using both numerical simulations and analytic calculations. 
For small values of $\rho_0$ (black, $\rho_0=0.10$), nodes are hardly activated even in the originating community; the activation essentially fails to propagate, regardless of $\mu$. 
By increasing $\rho_0$ (blue, $\rho_0=0.13$), one reaches a threshold where the contagion can spread to the whole originating community if $\mu$ is sufficiently small. 
However, when a critical value of $\mu$ is exceeded, the internal connectivity becomes insufficient to spread the contagion to the whole originating community. 
As the originating community is not saturated, the diffusion does not spread to the other community as well.
In this situation there is no overlap between the blue and red area in Fig.~\ref{fig:optimum_mu}. 

A larger value of $\rho_0$ (red, $\rho_0 =0.17$) finally allows the global
diffusion.  The range of values of $\mu$ that allows full activation in the
originating community is even further extended (fewer internal links are
needed), until a sufficient number of links can be \textit{spared} to induce
full activation in the second one.  If, however, the number of intra-community
links becomes too small (large $\mu$), the activation fails to spread in the
originating community and therefore it cannot be transmitted over the entire
network, despite the increased number of cross-community links.  The above
reflects in a finite, intermediate range of community strength that allows
global spreading. 

Even larger values of $\rho_0$ (red and magenta) simply extend the range of
$\mu$ for which the activation of the entire network is achieved.  When
$\rho_0$ becomes larger than the critical value for the transition in networks
without communities, increasing $\mu$ never blocks the local spreading, and
thus the global diffusion always happens as long as the network has enough
bridges.  Notice that $\rho_\infty$ is always larger for intermediate values of
$\mu$ with respect to the no-community case ($\mu=1/2$) and indeed full
activation can be obtained in an ample set of values of $\rho_0$ if $\mu$ is
properly chosen.  The smallest value of $\mu$ that allows full activation of
the second community is essentially independent of $\rho_0$, if $\rho_0$ is
sufficiently large: once the first community is fully active it is only a
matter of providing sufficient external links, therefore the precise value of
$\rho_0$ does not matter.  Specifically, using the TL formulation and the
present value of $\theta$, we obtained that $\mu_c \simeq 0.2175$ requires the
minimal amount of seeds compatible with global diffusion.  The value of $\mu$
for which the decay of $\rho_\infty$ sets in, instead, results from not having
sufficient internal links to achieve full activation of the originating
community given the initial seed. The value of $\mu$ depends therefore on
$\rho_0$. 

Although we here present results only for the case of random networks with two
communities and a specific value of $\theta$, our results are more general.  In
the supplementary material we provide evidence that our results are robust
under changes in the number of communities and assuming degree distributions
more general than that induced by the random arrangement of links described
above. Our results include experiments run on LFR benchmark
graphs~\cite{lancichinetti2008benchmark} that provision for a power law degree
distribution both for the degree and the size of multiple communities. It is
also worth stressing that both the MF and TL methods are flexible enough to
handle arbitrary (and community-specific) degree distributions, and arbitrary
inter-community connectivity patterns.  To adapt MF to this general case one
would need to replace (e.g., in the equation for $\rho_{\infty}^{A}$ in
Eq.~\ref{eq:meanfield2}) $p(k)$  with the specific degree distribution of
community A and $q^A$ with $\sum_{J \in \mathcal{C}} p_{AJ}\rho_{\infty}^{J}$,
where $\mathcal{C}$ is the set of communities and $p_{AJ}$ is the probability
that a link departing from a node in  $A$ ends in $J$.  In the supplementary
material, we also provide evidence that our results are qualitatively unchanged
by varying the system size $N$, the average degree $z$, and other parameters.
Finally, our results are also robust for changes in the threshold $\theta$ for
a pretty wide range of values (see supplementary material).

In summary, our analysis shows that there exists an optimal strength of community structure that facilitates global diffusion. 
We demonstrate that the presence of the right amount of community structure may, counterintuitively, enhance the diffusion of information rather than hinder it.  
A tight community, with its high level of internal connectivity, can act as an incubator for the localized information diffusion and help to achieve a critical mass. 
Information can then spread outside the community effectively as long as sufficient external connectivity is guaranteed.  
Our results enrich the growing body of literature that stresses the influence of the community structure in a large number of processes, including epidemics, viral marketing, opinion formation, and information diffusion. Our findings can be generalized,
and offer insights to understand recent empirical observations, such as the counterintuitive behavior of information diffusion in clustered networks~\cite{centola2010spread}, or the strong link between viral memes and the community structures in Twitter~\cite{weng2013virality}.
Further work is needed to understand how our observations hold if different mechanisms of transmission are considered, or a richer and more complex organization of communities is assumed.

\begin{acknowledgments}

We thank James P. Bagrow, Filippo Menczer, and Sergey Melnik for helpful
discussions and suggestions. During the review process, we have been made aware
of a similar result independently observed by S. Melnik \emph{et
al}~\cite{melnik-lfr-2014}. AF acknowledges NSF (grant CCF-1101743) and the
McDonnell Foundation. AF and EF acknowledge DARPA (grant W911NF-12-1-0037).  YA
acknowledges support from Microsoft Research. 

\end{acknowledgments}


\section{Supplementary Material}

\subsection{Introduction} 

The study of diffusion and spreading processes---such as epidemic spreading,
fads, diffusion of innovations, and viral memes---is a fundamental topic in a
number of disciplines, including mathematical, physical and social sciences,
communication, marketing and economics. In this paper, we investigate how
information diffusion is affected by network community structure using the
linear threshold model. We simulate information diffusion as following: at time
$t=0$, a fraction $\rho_0$ of randomly selected agents, `seeds', in one of the
communities is initialized in the active state. At each time step, agents'
states are updated synchronously if the fraction of its neighbors who are
already in the active state will be greater than the adoption threshold ($\theta$).
Using analytical approaches and numerical simulations, we demonstrate the
existence of nontrivial optimal modularity, where global cascades require the
minimal number of early adopters.

In the present supplementary material we corroborate broaden the scope of our findings by reporting
additional results on a wide range of configurations and parameter spaces.
Specifically, we vary the threshold parameter $\theta$, the average degree (and
therefore clustering coefficient) $z$, the network size $N$, the number of communities,
the degree distribution, and the community size distribution. 


\subsection{Results} 

\subsubsection{Average degree and clustering coefficient} 

\begin{figure}
\centering
\includegraphics[width=1.0\columnwidth, clip=true, trim=0 10 0 8]{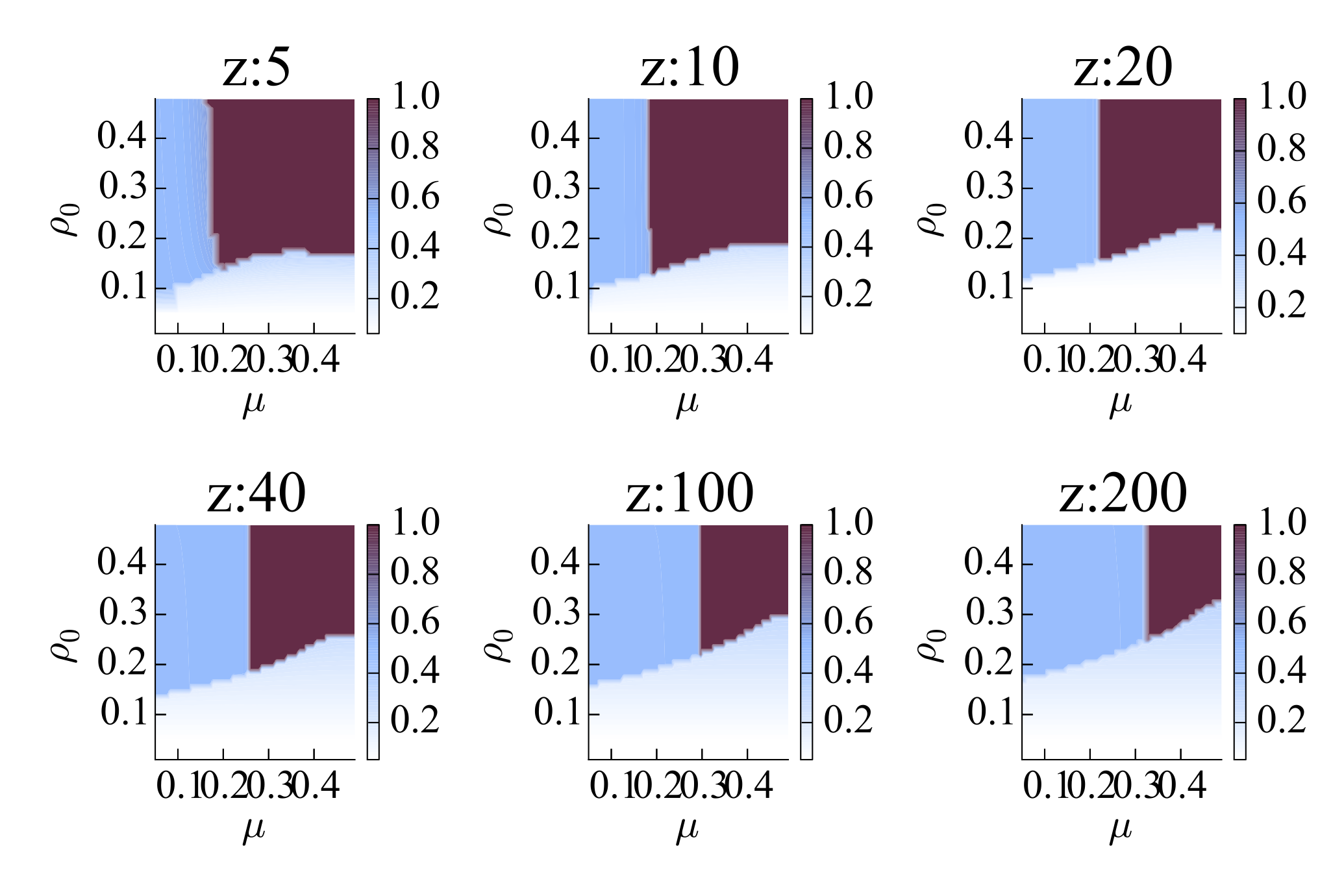}
\caption{The phase diagram of the threshold model using TL in the presence of community structures for different $z$ and $\theta=0.4$.}
\label{fig:fig_z_new}
\end{figure}

\begin{figure}
\centering
\includegraphics[width=1\columnwidth, clip=true, trim=0 10 0 8]{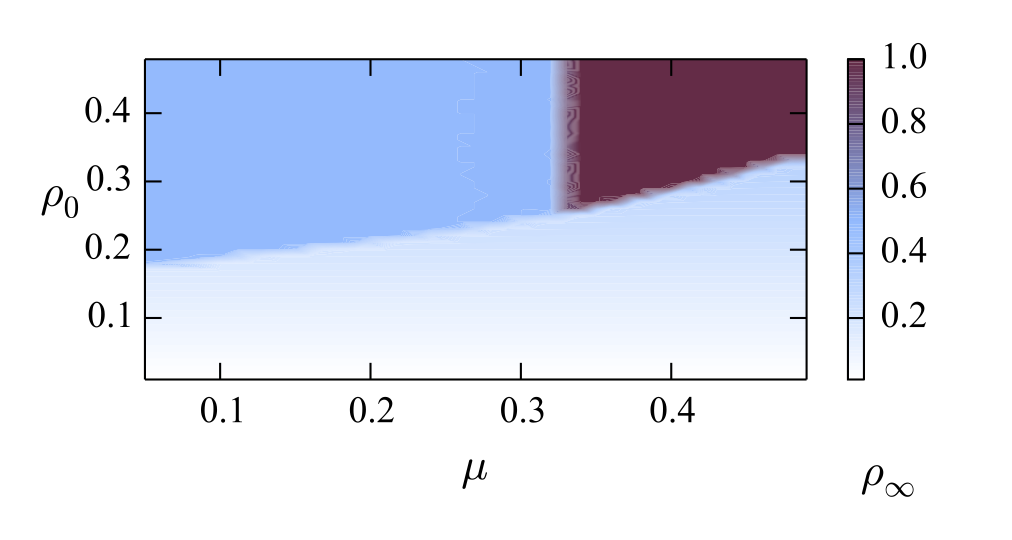}

\caption{(a) The phase diagram of the threshold model using simulation in the presence
of community structure with $N = 1024$, $z=200$, and $\theta=0.4$.  
 \label{fig:all_tl_mf}}

\end{figure}

Figure~\ref{fig:fig_z_new} shows results with various values of average degree.
The change in average degree does not change the behavior of
the dynamics. As the average degree increases, the optimal value of $\mu$ also
increases. 

Figure~\ref{fig:all_tl_mf} demonstrates that strong clustering hardly changes
the qualitative behavior of the system. In this simulation clustering coefficient for $\mu=0.23$ is
$0.226$ and for $\mu=0.01$ is $0.38$. 


\subsubsection{Disassortative (bipartite) mixing}  

\begin{figure}

\centering
\includegraphics[width=\columnwidth, clip=true, trim=0 10 0 8]{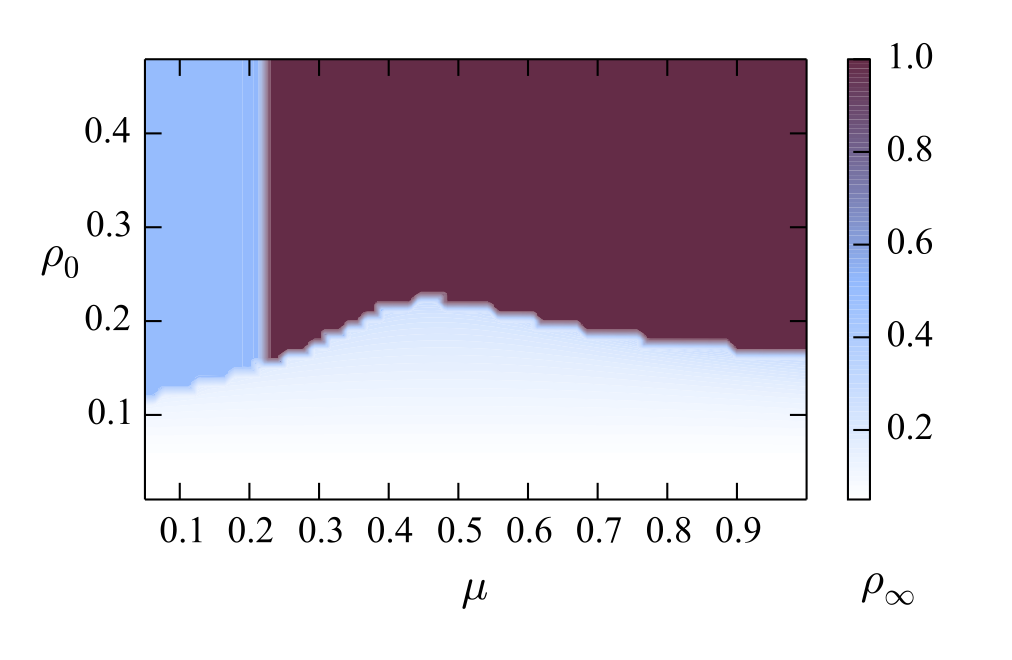}

\caption{The phase diagram of threshold model using TL  in the presence of
community structure, with $z=20$ and $\theta=0.4$.  Figure shows three regions
in which  $\mu< 0.5$ (assortative; modular),  $\mu = 0.5$ (random) and  $\mu> 0.5$
(dissasortative; bipartite).  } \label{fig:all_mu}

\end{figure}

Figure~\ref{fig:all_mu} shows the phase diagram across the whole parameter
range of $\mu$. An interesting pattern emerges: the minimum $\rho_0$ value that
gives rise to a  global cascade increases until around $\mu =
0.45$ and decreases thereafter, while the optimal mixing parameter remains
  around 0.25. The global cascade for large values of  $\mu$ is explained by a
qualitatively different dynamical scenario: rather than developing in the
originating community and then spreading in the second, adoption spreading
alternates longitudinally between the two communities thanks to the increased
level of connectivity among them. 

\subsubsection{Network size} 

\begin{figure}
\centering
\includegraphics[width=\columnwidth, clip=true, trim=0 10 0 8]{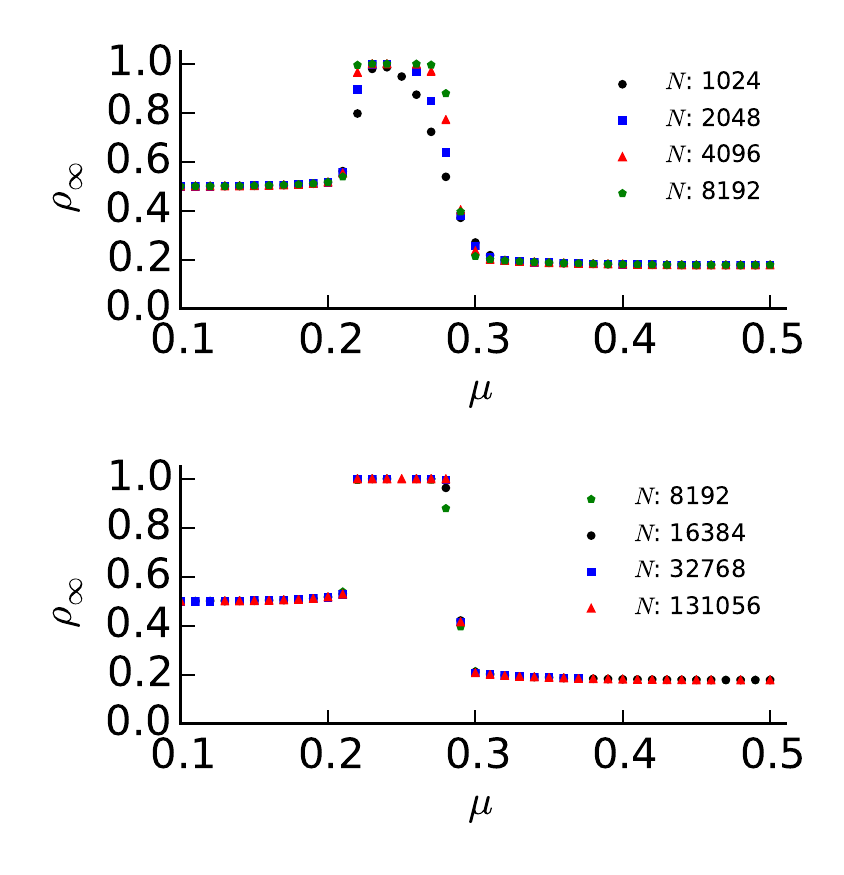}
\caption{The phase diagram of the threshold model, derived through simulations, in the presence of community structure for different network sizes, with $z=20$ and $\theta=0.4$.}
\label{fig:fig_size}
\end{figure}

Figure~\ref{fig:fig_size} shows that the size of networks does not affect our
results. 

\subsubsection{Number of communities} 

\begin{figure}
\centering
\includegraphics[width=\columnwidth, clip=true, trim=0 10 0 8]{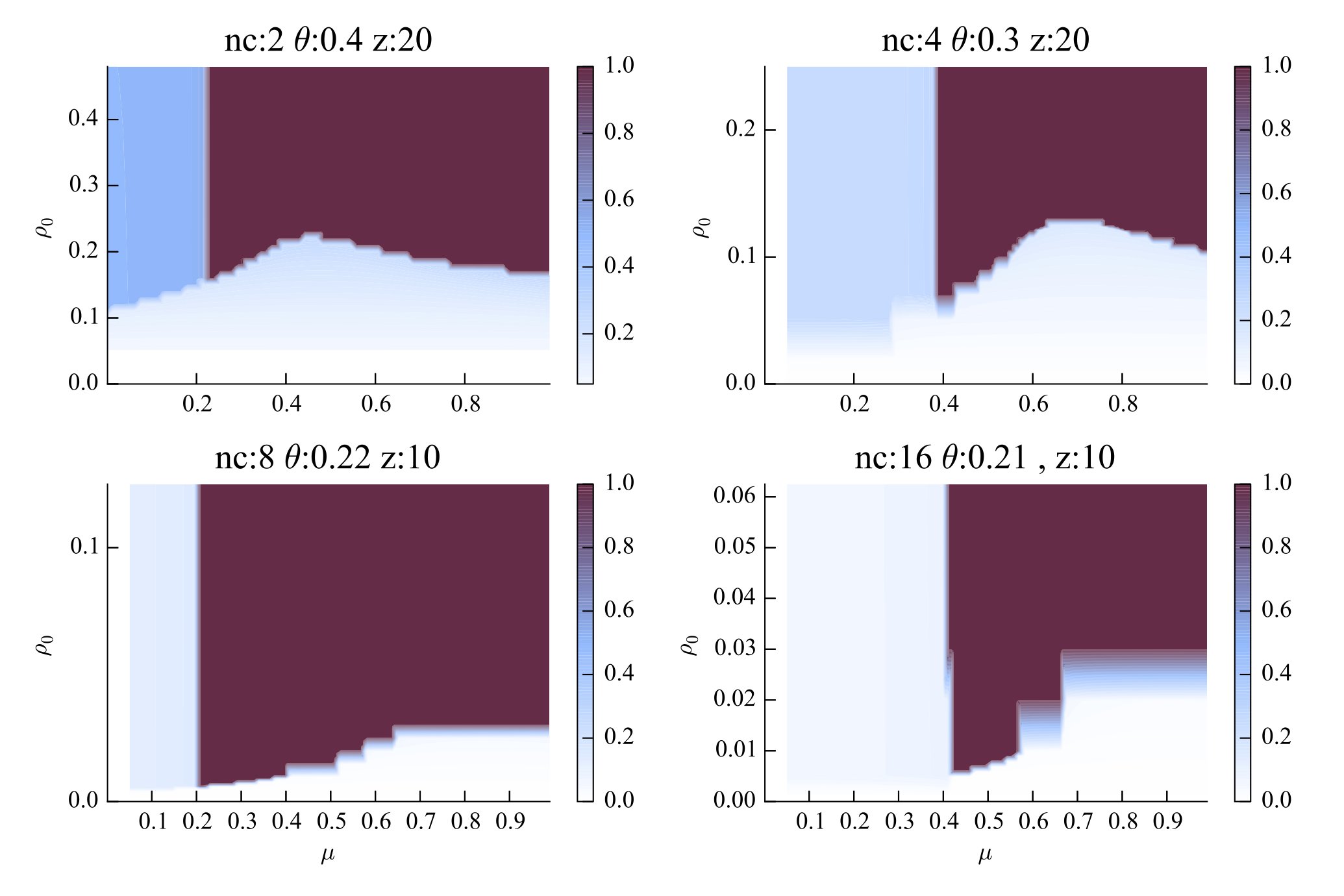}
\caption{The phase diagram of threshold model in the TL approximation with different number of communities.} \label{fig:fig_commu}

\end{figure}

\begin{figure}
\centering
\includegraphics[width=\columnwidth, clip=true, trim=0 10 0 8]{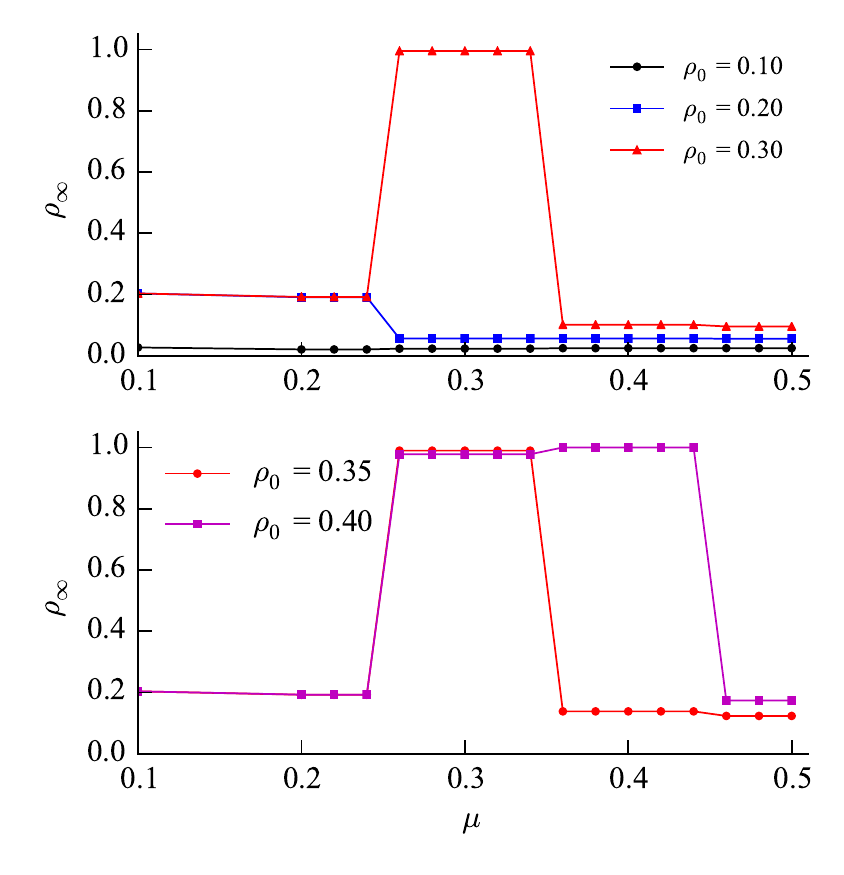}

\caption{The behavior of threshold model in the presence of community
structures generated by LFR benchmark, with $N=25000$, $z=10$, $t_1=2.5$
(degree exponent), $t_2=1.5$ (community size exponent), $k_{max}=30$ and
$\theta=0.3$.  LFR benchmark generates more \emph{realistic} networks with
community structures. The degree distribution may have a power-law distribution
(with exponent $t_1$ and degree cutoff $k_{max}$). The size of the communities
may also follow a power-law distribution (with exponent $t_2$).
\label{fig:fig_lfr}}

\end{figure}

We have repeated our calculation on networks with varying number of
communities. We considered both the case in which intra and extra-commuity
connectivity is random (analogously to the case discussed in the manuscript)
and the more general case of networks generated by the so-called LFR benchmark
graphs~\cite{lancichinetti2008benchmark}. The LFR benchmark framework is
commonly used in testing community detection algorithms because it can generate
more realistic---with heterogeneous degree and community size
distribution---networks. Figure~\ref{fig:fig_commu} demonstrates that the
optimal modularity behavior persist when the number of communities is larger
than two (see also~\cite{melnik-lfr-2014} for a similar observation).
Figure~\ref{fig:fig_lfr} shows that it is possible to observe the same
qualitative behaviors in LFR networks in which degree distribution and
community size distribution are generated from power-law functions. 

It is worth to emphasize that increasing the number of community can affect
other parameters such as adoption threshold. More community requires smaller
adoption threshold to complete the cascade since increases in  number of
community leads to decrement in the number of bridges among communities. 

\subsubsection{Adoption threshold  } 

\begin{figure}
\includegraphics[width=\columnwidth, clip=true, trim=0 10 0 8]{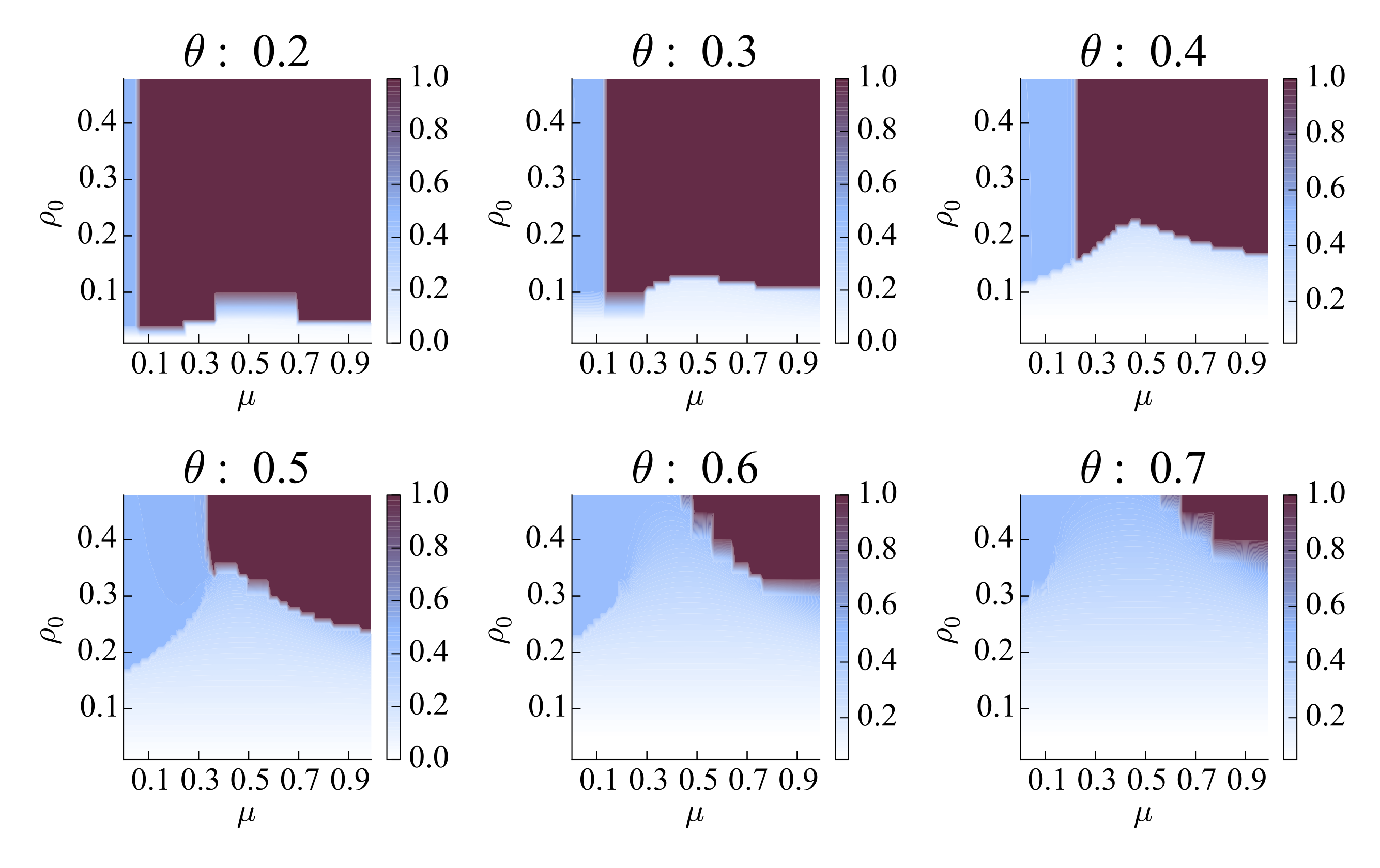}
\caption{The phase diagram of threshold model using TL  in the presence of community structures with  $z=20$ for different values of $\theta$. 
} \label{fig:all_theta}
\end{figure}

\begin{figure} 

\includegraphics[width=\columnwidth, clip=true, trim=0 10 0 8]{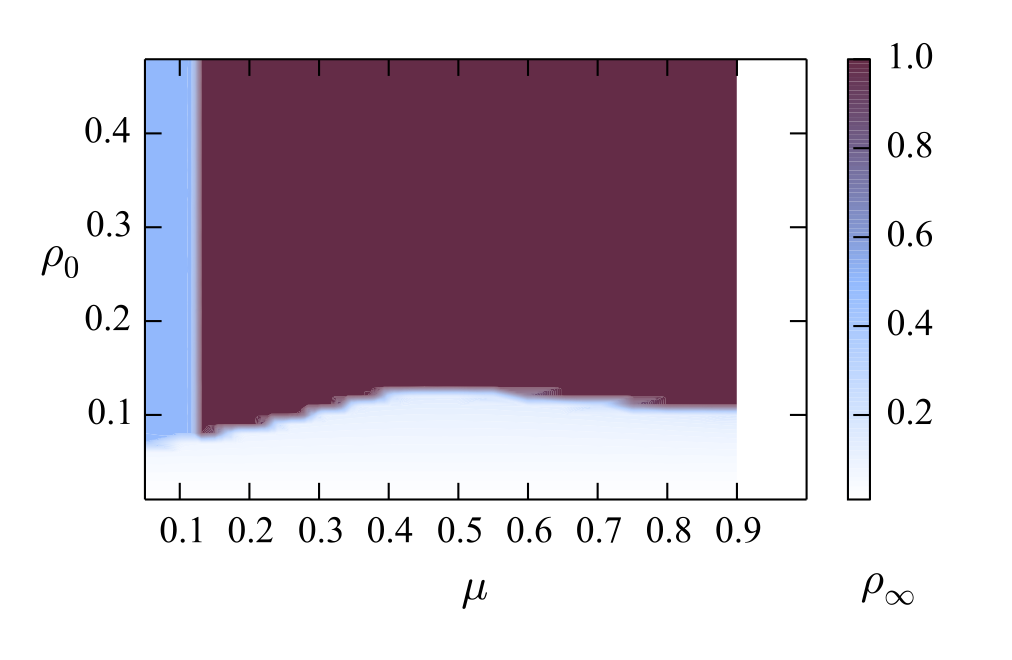} 
\includegraphics[width=\columnwidth, clip=true, trim=0 10 0 8]{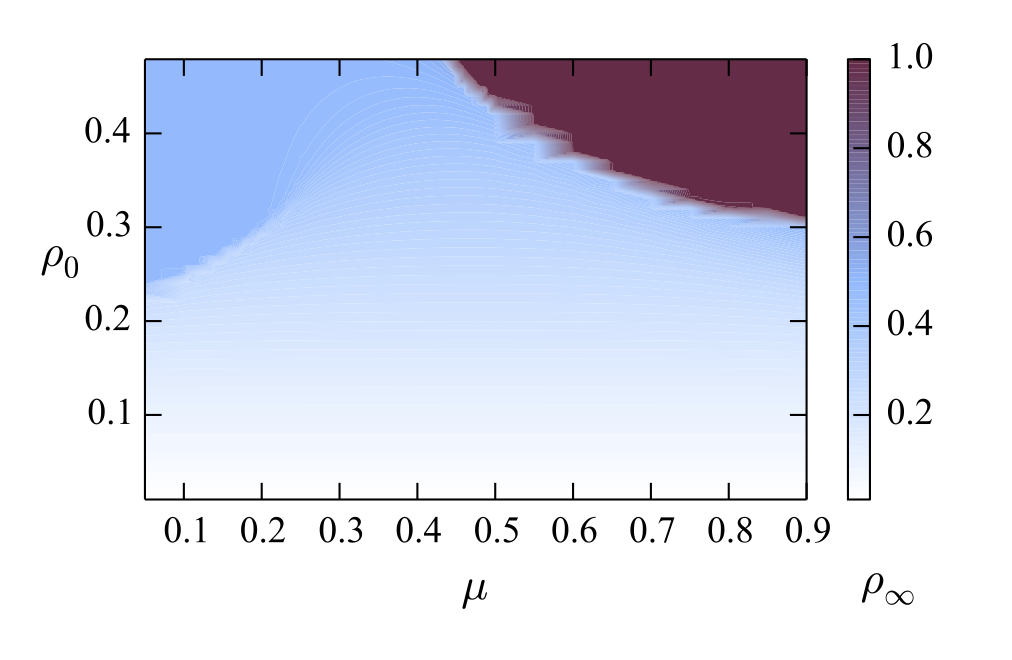} 

\caption{The phase diagram of threshold model using simulation in the presence
of community structure with $N = 8192$, $z=20$, and (Left) $\theta=0.3$
(Right) $\theta=0.6$. These results confirm the analytic solution shown in
the previous figure.  } \label{fig:theta_03}

\end{figure}

Adoption threshold $\theta$ controls how easy for the contagion to spread.
Figure~\ref{fig:all_theta} (TL approximation) and~\ref{fig:theta_03}
(simulation) shows the phase diagrams for various values of $\theta$. They 
demonstrate that qualitatively similar behavior can be observed across a wide
range of threshold values which extends up to approximately $\theta=0.5$.
Interestingly, if $\theta>0.5$, the behavior changes qualitatively. When the
threshold is very large, higher value of $\mu$ allows global cascade earlier.
The initial cascade happens only when the seed size is large and modularity is
very high ($\mu \simeq 0$) or very low ($\mu \simeq 1$). When the modularity is
very high, the cascade does not extend beyond the originating community. By
contrast, strong disassortative structure allows a cascade to spread in a
longitudinally alternating way between the two communities.  

\bibliographystyle{unsrt}
\bibliography{refs}

\end{document}